\documentclass[fleqn,usenatbib]{mnras}

\begingroup\lccode`~=`^
\lowercase{\endgroup\def~}#1{^{\scriptscriptstyle#1}}

\AtBeginDocument{\mathcode`^="8000 \catcode`^=12 }

\begingroup\lccode`~=`_
\lowercase{\endgroup\def~}#1{_{\scriptscriptstyle#1}}

\AtBeginDocument{\mathcode`_="8000 \catcode`_=12 }

\usepackage[font=small,labelfont=sc, labelsep=period, format=hang, singlelinecheck=false, justification=raggedright, textformat=period]{caption}

\usepackage[T1]{fontenc}
\DeclareRobustCommand{\VAN}[3]{#2}
\let\VANthebibliography\thebibliography
\def\thebibliography{\DeclareRobustCommand{\VAN}[3]{##3}\VANthebibliography}

\usepackage{graphicx}	
\usepackage[tbtags]{amsmath}	
\usepackage{tabularx,booktabs}

\usepackage{threeparttable}
\usepackage[normalem]{ulem}

\usepackage{amssymb}	
\usepackage{arydshln}

\usepackage[labelformat=simple]{subcaption}

\usepackage{cleveref}
\usepackage{mwe}

\usepackage{titlesec}
\titlespacing*{\subsection}{0pt}{0.8\baselineskip}{0.2\baselineskip}
\titlespacing*{\section}{0pt}{2\baselineskip}{0.5\baselineskip}
\usepackage{newtxtext,newtxmath}

\usepackage{hyperref}
\hypersetup{colorlinks,%
citecolor=black,%
filecolor=black,%
linkcolor=black,%
urlcolor=black
}

\title[Satellite Tidal Stalling]{Tidal Dissipation in Satellites Prevents Hill Sphere Escape}

\author[Kisare \& Fabrycky]{
A. M. Kisare\thanks{E-mail: akisare@uchicago.edu}, 
D. C. Fabrycky
\\
Dept. of Astronomy \& Astrophysics, University of Chicago, 5640 S. Ellis Ave., Chicago, IL 60637
}

\date{Accepted XXX. Received YYY; in original form 31 August 2023}

\pubyear{2023}

\begin{document}
\newcommand{\rH}{r_{\rm H}}
\label{firstpage}
\pagerange{\pageref{firstpage}--\pageref{lastpage}}
\maketitle

\begin{abstract}
The transit method is a promising means to detect exomoons, but few candidates have been identified. For planets close to their stars, the dynamical interaction between a satellite's orbit and the star must be important in their evolution. Satellites beyond synchronous orbit spiral out due to the tide raised on their planet, and it has been assumed that they would likely escape the Hill sphere. Here we follow the evolution with a three-body code that accounts for tidal dissipation within both the planet and the satellite. We show that tidal dissipation in satellites often keeps them bound to their planet, making exomoons more observable than previously thought. The probability of escape depends on the ratio of tidal quality factors of the planet and satellite; when this ratio exceeds $0.5$, escape is usually avoided. Instead, the satellite moves to an equilibrium in which the spin angular momentum of the planet is not transferred into the orbit of the satellite, but is transferred into the orbit of the planet itself. While the planet continues spinning faster than the satellite orbits, the satellite maintains a semi-major axis of approximately $0.41$ Hill radii. These states are accompanied with modest satellite eccentricity near $0.1$ and are found to be stable over long timescales.
\end{abstract}




\section{INTRODUCTION}
Satellites of planets in the Solar System are common, and the giant planets all host extensive systems of satellites. Their existence, composition, and orbital states have all contributed to theories of the origin of the planets themselves \citep{1999Peale}. With the discovery of more than 5000 exoplanets, it is natural to ask whether those bodies have so-called exomoons. 

The most straightforward way to detect exomoons to date involves both the planet and the exomoon transiting in front of the star \citep{1999Sartoretti}. Even if the moon itself is not seen transiting, the timing variations it induces on the planet may be seen \citep{2009Kippinga}. The first precise transit curves of the first transiting planet were thus searched for exomoons \citep{2001Brown}.

However, Only a few candidates have been proposed (Kepler-1625, \citealt{2018Teachey}; Kepler-1708, \citealt{2022Kipping}). The community has pushed back against the first of those detections \citep{2019Heller, 2019Kreidberg, 2018Rodenbeck} despite refined analyses \citep{2018Teacheyb, 2020Teachey}. Other groups proposing candidates based on transit timing \citep{2021Fox}, have also received constructive criticism \citep{2020Kipping}.

With a lack of any definitive detections so far, theory becomes more important, both to reexamine what we really expect for the frequency of exomoons, and perhaps to provide some guidance on where the search could be most fruitful going forward. 

The most immediate reason we expect exomoons is that moons are nearly ubiquitous in the Solar System, and there are theoretical reasons to expect moons around exoplanets may be even more detectable than straight analogues to the Solar System satellites \citep{2018Hamers,2020Moraes}. Since satellite formation has numerous { pathways}, it is hard to develop confident predictions. 

Many have done tidal calculations for moons in the Solar System and hypothetical moons around exoplanets. Once a moon has formed, \cite{2002Barnes} calculated the tidal evolution, showing that planets which are tidally locked with their stars will cause their satellites to decay, but rapidly-spinning planets could push their satellites out of the Hill sphere. In fact, \cite{1973Counselman} had already speculated that this type of evolution may have removed primordial moons from Mercury and Venus. Both \cite{2012Sasaki} and \cite{2018Piro} also took into account the slowing of { a planet's rotation} by the tide raised by a moon. \cite{2009Cassidy} looked at the problem even closer, to see how the tidal response of the moon might change the dynamics. We pick up the theoretical story there.

The Hill sphere is the region around a planet where the gravity of a planet dominates the motion of test particles (and real moons, to a good approximation). Its size is $\rH = a_{\rm p} (m_{\rm p}/3m_*)^{1/3}$, where $a_{\rm p}$ is the planet's semi-major axis and $m_{\rm p}$ and $m_*$ are the mass of the planet and star, respectively. At this distance from the planet lie the Lagrange $L_1$ and $L_2$ equilibrium points. These are unstable equilibria though, and a natural satellite placed at them would eventually fall into either astrocentric or planetocentric orbits. This radius limits the motion of a satellite, but stable motion does not usually extend to that distance. Rather, prograde orbits have $0.5 r_{\mathrm{H}}$ as a typical maximum distance of stability \citep{2008ShenTremaine}, and orbits initialised further are able to escape beyond $L_1$ and $L_2$. The evection resonance, in which the  apsidal period of the satellite equals the planet's orbit around the star, is thought to be the mechanism that finally disrupts a satellite orbit \citep{2017Grishin}. In a rotating reference frame, the fictitious force known as the Coriolis force enhances the binding of retrograde satellites and destabilizes prograde ones \citep{1980Innanen}. Though contained within a zero-velocity curve, stable orbits can quickly transition between circular and eccentric \citep{1991Hamilton}. These works did not include tides, so we suspected these orbits with fluctuating eccentricities may evolve due to dissipation. We set out to find the outcome.

In the following sections, we describe our numerical methods (\S~\ref{sec:methods}), display the results of a suite of n-body simulations (\S~\ref{sec:results}), derive analytically the stability conditions (\S~\ref{sec:analysis}), and explain how the results affect the big picture of exomoon evolution and even detection (\S~\ref{sec:conclusions}).

\section{METHODS} \label{sec:methods}
 \begin{figure}
	\includegraphics[width=\columnwidth]{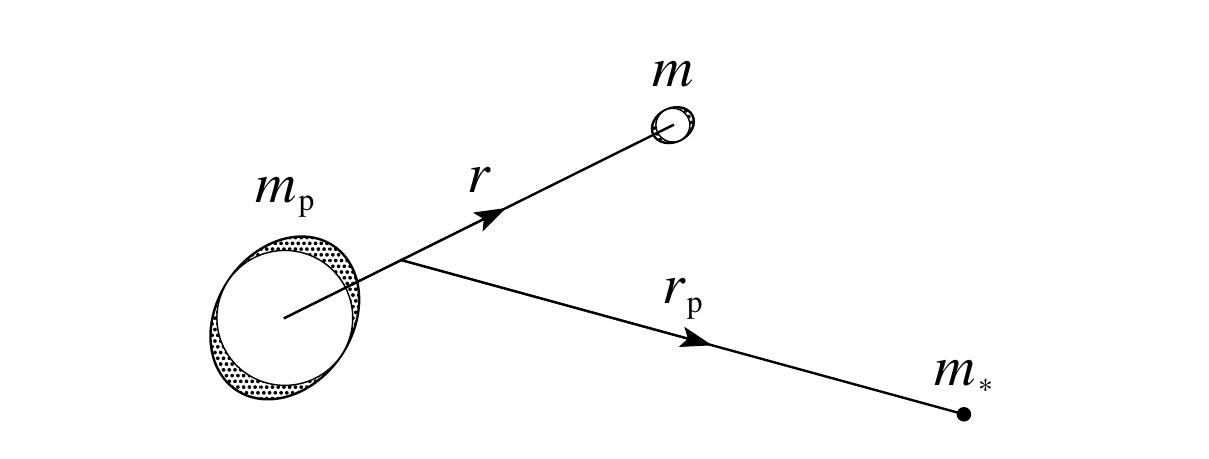}
    \caption{Model components. Tidally distorted and shifted planet and satellite, with a point-mass star referred to the centre of mass of the inner system. For the purposes of the n-body code, the center of $m_{\rm p}$ is treated as the origin. Relative sizes and distances are not to scale.}
    \label{fig:model}
\end{figure}
Our method involves direct N-body simulations, using the ``direct code'' of \cite{2002MardlingLin}, described in their Section 2. { The direct code uses a Bulirsch-Stoer algorithm with four integration steps per circular orbit of the innermost body.} {~\cite{2017Zollinger} noted that Mardling \& Lin's equation (7) contains a minor error; the term $m_3\beta_{34}$ within the first set of brackets should be changed to $m_4\beta_{34}$. However, the equations of motion in the code provided to us by Mardling do not contain this error.} The hierarchical coordinate system is illustrated by Figure~\ref{fig:model}.

\subsection{Variables} \label{subsec:variables}
Table~\ref{tab:params4} defines many variables used throughout this paper. { Quantities with a subscript ``$\rm p$'' reference the planet}, starred quantities reference the stellar body, and quantities without either generally refer to the satellite unless otherwise clarified. We describe the orbital elements of the inner pair as those of the satellite around the planet, and the orbital elements of the outer orbit as those of the planet around the star. However, the coordinates system used by the code is in the frame centred on the planet. When plotting results, rather than using Keplerian elements, which were found to be very poor at qualitatively describing the orbits this far out in the Hill sphere, we use $a = (r_{\rm max}+r_{\rm min})/2$ and $e=(r_{\rm max}-r_{\rm min})/(2a)$. $r_{\rm min}$ and $r_{\rm max}$ were determined by splitting the positional data into 625 intervals with 32 data points per interval, calculating the minimum and maximum radial position in each interval and then interpolating between these values using a piece-wise cubic Hermite interpolating polynomial. { Additionally, the satellite is defined as having escaped once $r>\rH$}.

We limit our explorations to prograde, coplanar, Moon-like satellites, with zero-obliquity, Earth-like planets. Previous work has used an analytical approach \citep{2009Cassidy} to determine how a star perturbs a tidally-evolving satellite. Our numerical approach is able to see new effects, even though the model is kept simple. The planet was placed in a low-eccentricity orbit, also for simplicity. The parameters we varied were the $Q$-factors of the planet and satellite. The $Q$-factor controls the strength of tidal dissipation in a body and depends on its interior structure and composition \citep{1966Goldreich, 2009Greenberg}. The influence that the $Q$-factor has on the stability of satellites close to the Hill sphere is therefore important in predicting the properties of any exomoon systems that may be discovered.

\subsection{Normalization} \label{subsec:norms}
The semi-major axis is normalized such that $\alpha = a/\rH$. Note that the magnitude of $\rH$ hardly changes in our simulations. We further define a normalized time $\tau = t/t_{\rm dyn}$ where $t_{\rm dyn}$ is the dynamical timescale:
\begin{equation}
    t_{\rm dyn} \equiv n/\dot{n} = \frac{2}{9n}  \frac{Q_{\rm p}}{k_{\rm p}} \frac{m_{\rm p}}{m} \bigg( \frac{a}{R_{\rm p}} \bigg)^5.
    \label{eq:tdyn}
\end{equation}
The value of $t_{\rm dyn}$ is the theoretical timescale for the satellite to migrate outwards to its present orbit due purely to tidal dissipation in the planet.\footnote{For the purposes of plotting results, $t_{\rm dyn}$ is computed using the initial values of $a$ and $n$.} Finally, the spin frequencies $\Omega$ and $\Omega_{\rm p}$ are normalized by the satellite's mean motion such that $s=\Omega/n$ and $s_{\rm p}=\Omega_{\rm p}/n$. For $s=1$, the satellite is rotating synchronously. For $s_{\rm p}=1$, the planet is rotating synchronously with the satellite.

\subsection{Momentum Conservation} \label{subsec:conservation}
The total angular momentum of the system is given by:
\begin{equation}
    \boldsymbol{J} = \mu_{\rm p}(\boldsymbol{r}_{\rm p} \times \boldsymbol{\dot{r}}_{\rm p}) + \mu (\boldsymbol{r} \times \boldsymbol{\dot{r}}) + I_{\rm p} \boldsymbol{\Omega}_{\rm p} + I \boldsymbol{\Omega}.  \label{eq:J}
\end{equation}
where $\mu_{\rm p} =  m_* (m_{\rm p} + m)  / (m_* + m_{\rm p} + m )$ and $\mu = m_{\rm p} m/(m_{\rm p} + m)$. The four terms are, in order, the orbital angular momentum of the planet around the star $\boldsymbol{L}_{\rm p}$, the orbital angular momentum of the satellite around the planet $\boldsymbol{L}$, the spin angular momentum of the planet $\boldsymbol{S}_{\rm p}$, and the spin angular momentum of the satellite $\boldsymbol{S}$. The total angular momentum $\boldsymbol{J}$ should be conserved. To test how well the code conserves angular momentum, we ran a simulation with $m_*=0$ for simplicity, and $Q_{\rm p}=Q=100$. Over a period of $5$~Myr, the fractional error in the total angular momentum $|\delta J|/J$ grew approximately linearly to ${\sim} 0.1\%$.  The proportion of orbital angular momentum $L/J$ grew from $50\%$ to $60\%$ at the expense of $S_{\rm p}/J$, a value much greater than the numerical error. The only component with a contribution smaller than the error was the spin angular momentum of the satellite $S/J \sim 0.001\%$. However, as $S$ is entirely coupled to $L$ and $S_{\rm p}$, its value is not compromised by the numerical error.

\section{NUMERICAL RESULTS} \label{sec:results}
\begin{figure*}
    \includegraphics[width=\textwidth]{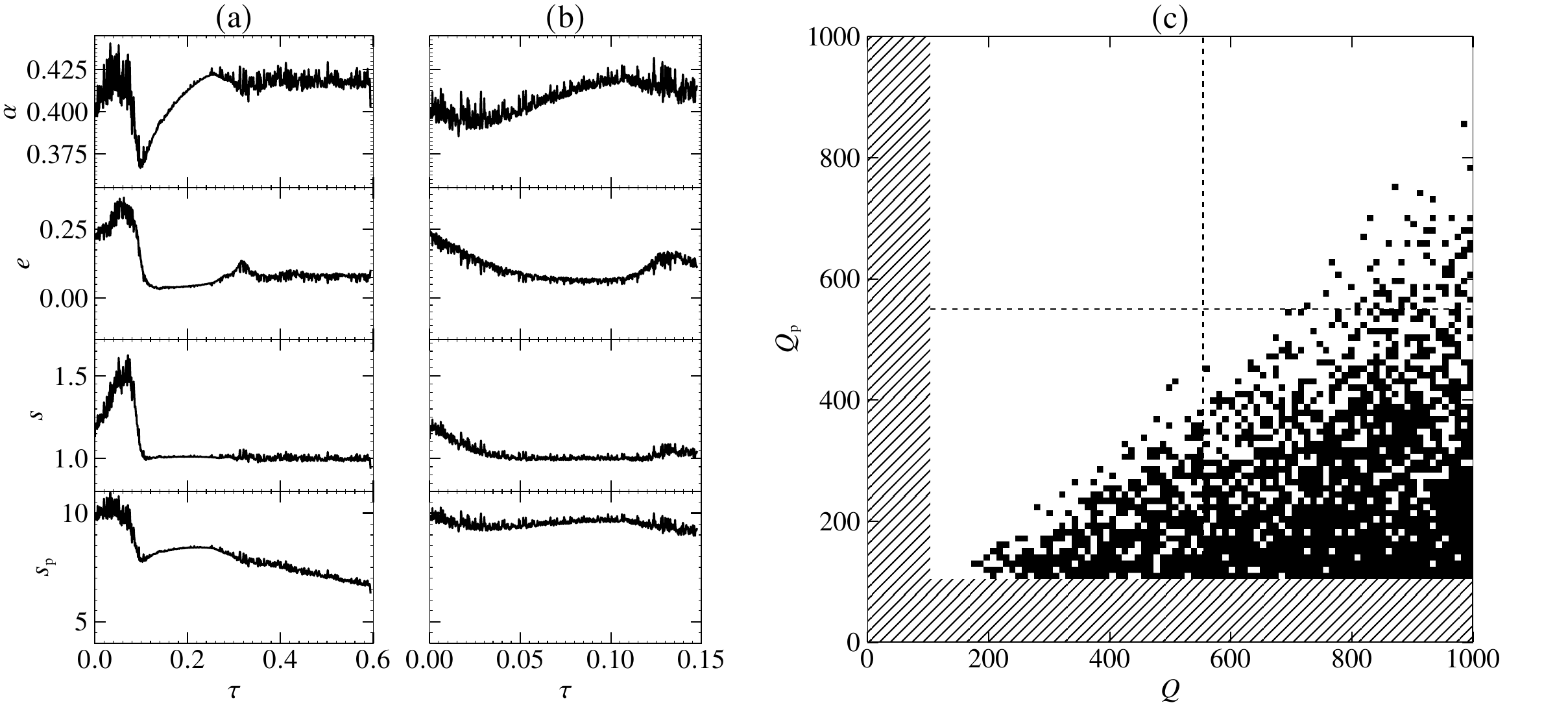}%
    \begin{subfigure}{0\textwidth}
        \phantomcaption
        \label{subfig:sim_a}
    \end{subfigure}%
    \begin{subfigure}{0\textwidth}
         \phantomcaption
         \label{subfig:sim_b}
    \end{subfigure}%
    \begin{subfigure}{0\textwidth}
         \phantomcaption
         \label{subfig:qplot}
    \end{subfigure}%
    
    \caption{Summary of numerical results: \subref*{subfig:sim_a} Simulation 1, with $Q_{\rm p}=Q=100$; \subref*{subfig:sim_b} Simulation 2, with $Q_{\rm p}~{\simeq}~403.48$ and $Q=100$; \subref*{subfig:qplot} set of simulations showing that both $Q_{\rm p}$ and $Q$ matter for whether the satellite stays bound to the planet. Each pixel represents a combination of $Q$ and $Q_{\rm p}$. All the white pixels had the satellite staying bound to the planet for the full duration of the simulation, while black pixels correspond to simulations where the satellite escaped. The hatched region indicates $Q$-factors that were not tested. The dotted line demarcates the four separate sets of simulation data that were stitched together. The upper left quadrant has the least variation in outcome, so a lower resolution of $0.02~{\rm px}/Q$ was used there, while the other three quadrants have a resolution of $(43/450)~{\rm px}/Q~{\simeq}~0.10~{\rm px}/Q$.}
    \label{fig:simplots}
\end{figure*}
In Figure~\ref{subfig:sim_a} we show an individual system evolution with $Q_{\rm p}=Q=100$, which we refer to as Simulation 1. The satellite approaches the Hill sphere until $\tau~{\simeq}~0.05$, after which the satellite is saved from escaping from the planet and falls back inwards. After $\tau~{\simeq}~0.3$, the satellite appears to approach an equilibrium, with $\alpha$ consistently fluctuating near $0.42$, moderate eccentricity and a stable spin frequency. The spin of the planet, however, continues decreasing for the entire duration of the simulation.

The angular momentum of the planetary spin is thus decreasing, yet the orbit of the satellite has stopped evolving; the angular momentum must be transferring into the planetary orbit, i.e. the orbit of the planet around the star is growing. { After equilibrium was achieved at $\tau~{\simeq}~0.3$, $L_{\rm p}/J$ grew linearly by approximately $0.2$~ppm from this point until the end of the simulation, while $S_{\rm p}/J$ decreased linearly by nearly the exact same amount, confirming that the planet's spin angular momentum is indeed being transferred into its own orbit.} We now experiment with two variables: the tidal quality factor of the planet $Q_{\rm p}$, and that of the satellite $Q$. 

In Figure~\ref{subfig:sim_b}, we let $Q_{\rm p}\simeq403.48$ and $Q=100$, which we call Simulation 2. In Simulation 1, which has a lower $Q_{\rm p}/Q$ than Simulation 2, the satellite very quickly approaches the Hill sphere, nearly escaping before having a crisis at $\tau~{\simeq}~0.05$, saving itself, falling back towards the planet, circularizing and reaching an equilibrium. In Simulation 2, the satellite never dramatically falls inwards before rebounding, and starts its evolution in a state similar to that of Simulation 1 at $\tau\simeq0.1$.

\begin{figure*}
    \includegraphics[width=\textwidth]{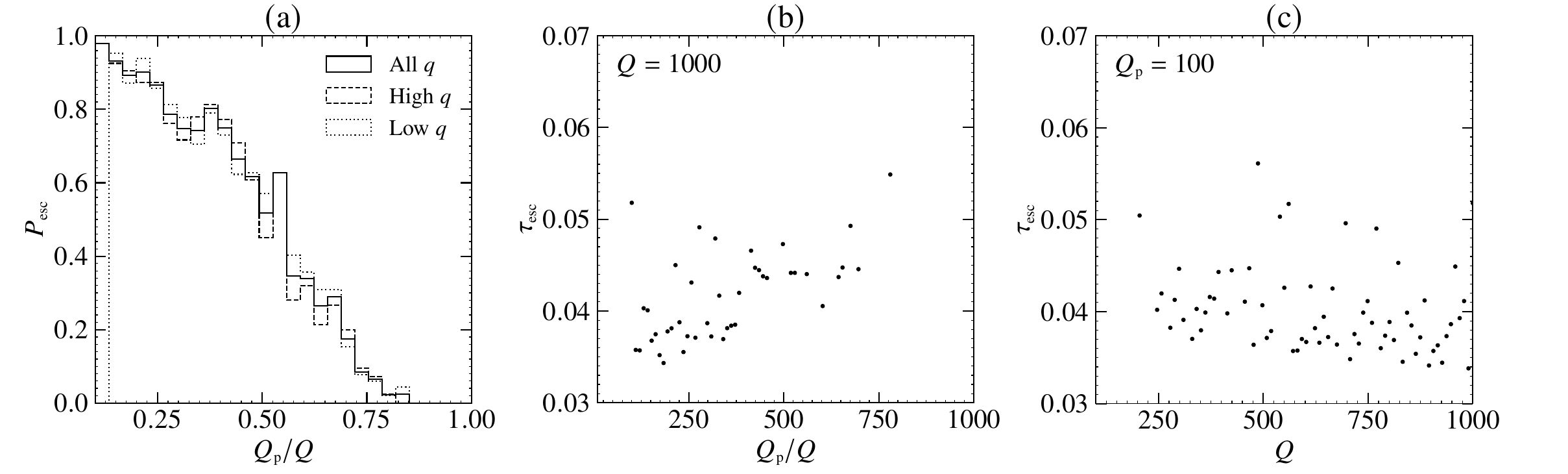}%
    \begin{subfigure}{0\textwidth}
        \phantomcaption
        \label{subfig:esc_a}
    \end{subfigure}%
    \begin{subfigure}{0\textwidth}
         \phantomcaption
         \label{subfig:esc_b}
    \end{subfigure}%
    \begin{subfigure}{0\textwidth}
         \phantomcaption
         \label{subfig:esc_c}
    \end{subfigure}%
    
    \caption{Statistics of escape: \subref*{subfig:esc_a} histogram of the probability of escape, versus the $Q$-ratio $Q_{\rm p}/Q$, with ``high $q$'' being the region where $743 < q < 1000$ and ``low $q$'' being the region where $q \leq 743$; \subref*{subfig:esc_b} relationship between escape time $\tau_{\rm esc}$ and $Q_{\rm p}$ for $Q=1000$; \subref*{subfig:esc_c} effect of $Q$ on escape time for $Q_{\rm p}=100$.}
    \label{fig:escapestat}
\end{figure*}

Running 5820 simulations, each with a unique combination of $Q_{\rm p}$ and $Q$, we find that escape occurs with complicated boundary (Figure~\ref{subfig:qplot}). We define $\boldsymbol{q}$ as the vector $(Q~Q_{\rm p})$ which has a slope $Q_{\rm p}/Q$ and a norm $q=|\boldsymbol{q}|$. From this set of simulation data, we organised each data point into 24 bins of slope ranging from 0.1 to 0.86. For each of these bins, the number of simulations that resulted in an escape of the satellite was divided by the total number of simulations in that bin, giving an escape probability $P_{\rm esc}$ in each bin. Points with $q > 1000$ were excluded, so as to help balance the number of data points in each bin. This is plotted as the ``all $q$'' histogram in Figure~\ref{subfig:esc_a}. Two more bins were then created, a ``high $q$'' bin where $743 < q < 1000$ and a ``low $q$'' bin where $q \leq 743$ . 743 was chosen as the boundary as this value equalises the number of data points in the high $q$ and low $q$ bins. The two additional histograms corresponding to these two bins are also plotted in Figure~\ref{subfig:esc_a}.

\begin{figure*}
    \includegraphics[width=\textwidth]{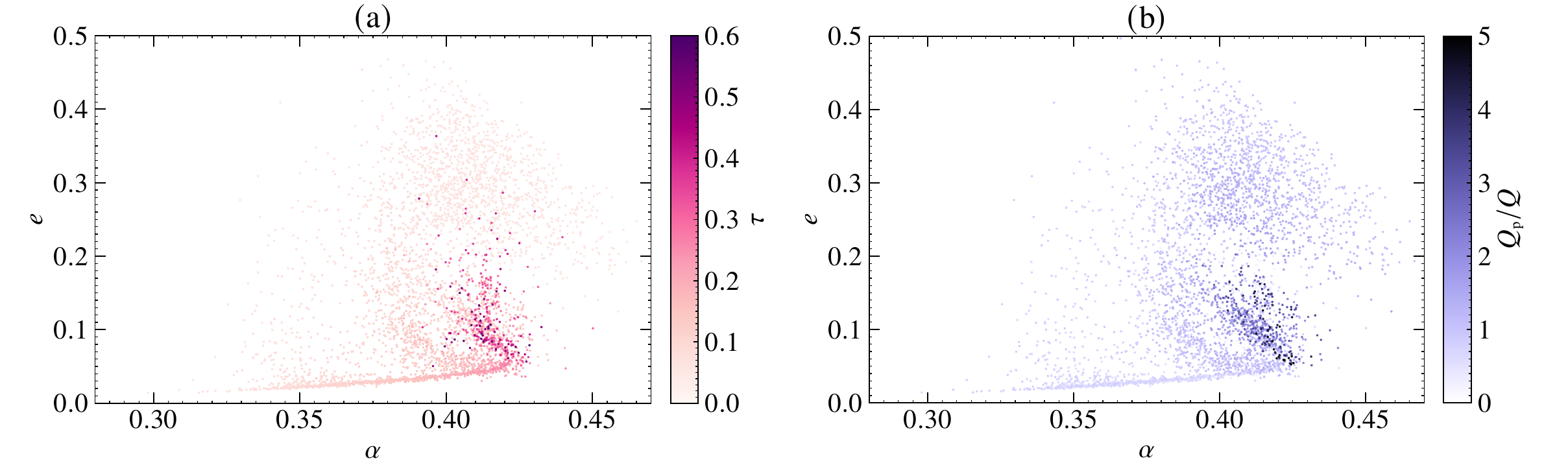}%
    \begin{subfigure}{0\textwidth}
        \phantomcaption
        \label{subfig:eq_a}
    \end{subfigure}%
    \begin{subfigure}{0\textwidth}
         \phantomcaption
         \label{subfig:eq_b}
    \end{subfigure}%
    
    \caption{ For satellites that did not escape, the final satellite orbits in $\alpha$-$e$ space: \subref*{subfig:eq_a} color-bar represents $\tau$, controlling how dynamically evolved the system is by the end of the simulation; \subref*{subfig:eq_b} color-bar represents $Q_{\rm p}/Q$ ratio: strong dissipation in the satellite relative to the planet is dark purple, whereas weak dissipation in the satellite relative to the planet is light purple.}
    \label{fig:equilibrium}
\end{figure*}

There is very little discrepancy between $P_{\rm esc}$ in the high-$q$ region and the low-$q$ region. The sharp drop in probability in the low-$q$ region around $Q_{\rm p}/Q~{\simeq}~0.1$ is a consequence of lower slopes being geometrically impossible. The consistency in the escape probability between the high and low regions suggests that escape is independent of $q$, but highly dependent on $Q_{\rm p}/Q$. Greater relative dissipation in the planet (low $Q_{\rm p}/Q$) yields higher probabilities of escape, while greater relative dissipation in the satellite (high $Q_{\rm p}/Q$) yields lower probabilities.

In Figure~\ref{subfig:esc_b}, we show that there exists a modest relationship between $\tau_{\rm esc}$ and $Q_{\rm p}$ for a given value of $Q$. $Q=1000$ was chosen as it maximizes the number of data points, but a similar relationship exists for all values of $Q$. Although there is much noise, escape generally occurs slightly quicker when $Q_{\rm p}$ is lower. This is interesting, as $\tau$ is already a time quantity normalized by the dynamical timescale $t_{\rm dyn}$ which from \eqref{eq:tdyn} is linear in $Q_{\rm p}$. Therefore, there is likely some weaker, higher-order dependence between $t_{\rm dyn}$  and $Q_{\rm p}$ which we have not accounted for. An opposite relation exists between $Q$ and $\tau_{\rm esc}$, shown in Figure~\ref{subfig:esc_c}. Stronger dissipation in the satellite (low $Q$) increases escape time, particularly at the boundary between escaping and remaining bound.

The statistical results at the end of all simulations for which the satellite did not escape is shown in Figure~\ref{fig:equilibrium}. { The value of $\tau$ in this plot is the normalized time at the end of the simulation. Higher $\tau$ represents systems that are more dynamically evolved.} This strongly suggests that, in Figure~\ref{subfig:eq_a}, there is a movement of orbits from the light-colored region to the dark-colored region centred around $e\simeq0.1$. From Figure~\ref{subfig:eq_b}, we find that highly evolved orbits can stabilise into this excited eccentricity if $Q_{\rm p}/Q>1$. Essentially, if the satellite does not escape, systems with strong dissipation in the satellite relative to the planet will stabilise the satellite into a slightly eccentric orbit.

The final state of the system in each of these simulations is with the planet still out of equilibrium; the planet is spinning down due to tides raised by the satellite on it. The Mardling code does not raise tides due to the star, but they would also act to spin down the planet. Future work will look at the even longer-term evolution of the satellite, once the planet has spun down to the orbital frequency of the satellite. We expect that from then on, the satellite cannot escape into an astrocentric orbit, because the total angular momentum of the planet-satellite system is too low to cause the satellite to reach the Hill sphere. Instead, there would be spin-orbit locking of both the planet and the satellite. Angular momentum would continue to drain at an even slower rate, now due mainly to the tidal bulge the star raises on the planet. This would couple to the satellite's orbit, now the dominant reservoir of angular momentum, which would very slowly decay.

\section{ANALYTIC RESULTS} \label{sec:analysis}
\begin{figure}
	\includegraphics[width=\columnwidth]{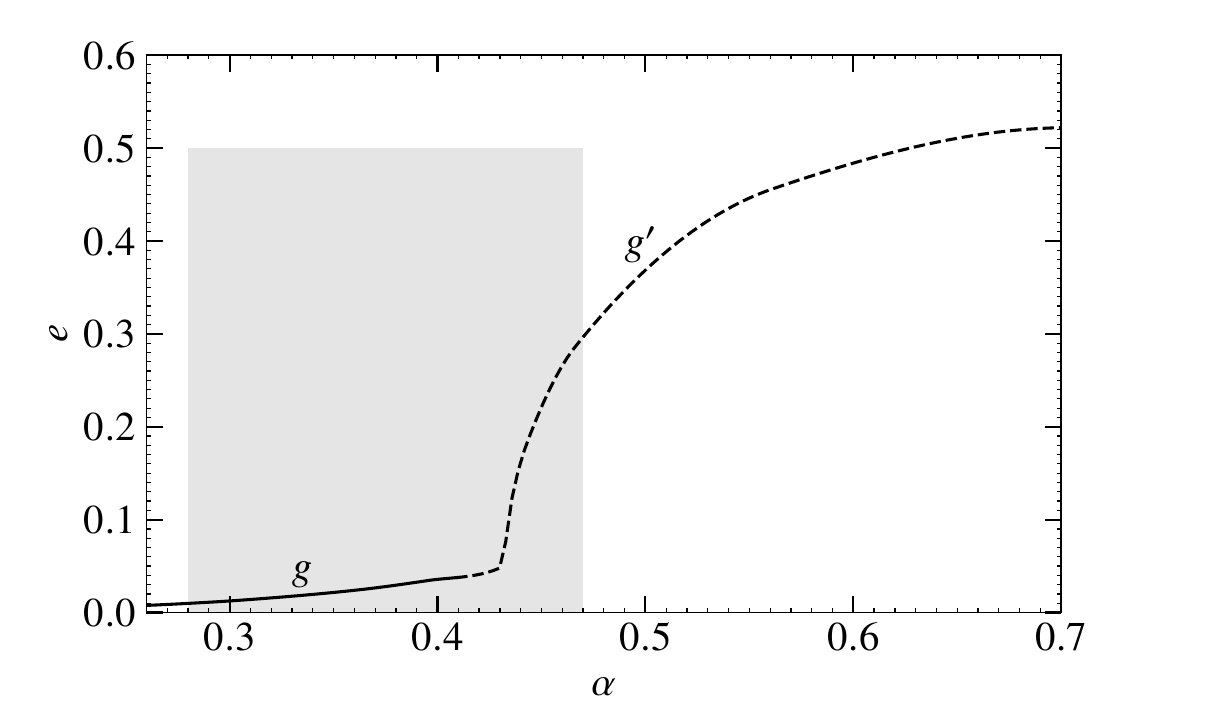}
    \caption{Periodic orbits within Hill's problem, modeling the point-mass three-body problem in which one of the three masses (the star) is much larger than the other two \citep{1969Henon}. The shaded region corresponds to the boundaries of the plots in Figure~\ref{fig:equilibrium}.}
    \label{fig:ae}
\end{figure}

Consider the the gravitational ``main problem'' of lunar theory \citep{1961Brouwer}, i.e. just three point masses and no rotational or tidal distortion, and no tidal dissipation. If the dominant mass and the barycentre of the two smaller masses follows a circular orbit, then periodic solutions of these two two smaller masses around each other exist \citep{1969Henon}. The two families of bound, periodic, prograde orbits are called $g$ and $g'$. Their $\alpha$ and $e$ are shown in Figure~\ref{fig:ae}. Family $g$ has low eccentricity, and its nearly-elliptical orbit is a centred ellipse, with long axis forming a right angle with the direction to the star. That is, it makes two radial oscillations for every azimuthal orbit. There is a maximum distance, $\alpha~{\simeq}~0.41$, beyond which no similarly shaped orbits are stable. Instead, family $g'$ curves continue to higher $\alpha$ at larger eccentricity,\footnote{Similarly in analyzing the truncation radius of circumplanetary disks, \cite{2011Martin} quoted $0.41 \rH$ as the point at which stream-lines would cross via further ones becoming eccentric, in Hill's problem.} and the figure is no longer centred on the planet but has its long axis pointing either towards or away from the star. That family also ends at $\alpha~{\simeq}~0.70$. 

\begin{table}
    \renewcommand*{\thefootnote}{\alph{footnote}}
	\caption{Variable Definitions and their Initial Values}
	\label{tab:params4}
	\begin{threeparttable}
    	\begin{tabularx}{\columnwidth}{@{}l @{}l *1{>{\centering\arraybackslash}X}@{}}
    	    \toprule
    	    \toprule
    		\multicolumn{2}{>{\centering\arraybackslash}c}{Variable and Definition} & Value\\
    	    \midrule
    		$a$ \qquad \qquad & Semi-major axis of satellite & {$0.404\rH$} \\\hdashline
                $a_{\rm p}$ & Semi-major axis of planet & {$0.18~\rm au$} \\ \hdashline
                $e$ & Eccentricity of satellite &{0.05}\\\hdashline
    		$e_{\rm p}$ & Eccentricity of planet & {0}\\\hdashline
    		$\varpi$ & Longitude of pericentre of satellite & {$0^\circ$}\\\hdashline
    		$\varpi_{\rm p}$ & Longitude of pericentre of planet & {$0^\circ$}\\\hdashline
    		$R$ & Radius of satellite & {$0.279R_{\oplus}$}\\\hdashline
    		$R_{\rm p}$ & Radius of planet & {$1R_{\oplus}$}\\\hdashline
    		$\Omega$ & Spin frequency of satellite\tnote{$\star$} &{$87.0^\circ/\rm day$}\\\hdashline
    		$\Omega_{\rm p}$ & Spin frequency of planet &{$864^\circ/\rm day$}\\\hdashline
    		$k$ & Love number of satellite & {0.8}\\\hdashline
    		$k_{\rm p}$ & Love number of planet & {0.6}\\\hdashline
    		$\tilde{I}$ & Moment of inertia coefficient of satellite & {0.35}\\\hdashline
    		$\tilde{I}_{\rm p}$ & Moment of inertia coefficient of planet & {0.3}\\\hdashline
    		$m$ & Mass of satellite & {$0.0123M_\oplus$}\\\hdashline
    		$m_{\rm p}$ & Mass of planet & {$1M_\oplus$}\\\hdashline
    		$m_*$ & Mass of star  & $1M_{\odot}$ \\\hdashline
    		$Q$ & Tidal quality factor of satellite\tnote{$\dagger$} & 100-1000 \\\hdashline
    		$Q_{\rm p}$ & Tidal quality factor of planet\tnote{$\dagger$} & 100-1000 \\
    		\bottomrule
    	\end{tabularx}
    	\begin{tablenotes}\footnotesize 
                \item[$\star$] Synchronised 1:1 with the mean motion of the satellite.   	    
                \item[$\dagger$] See \S~\ref{sec:results} for the particular $Q$-factors used in Simulation 1 and 2. A total of 5820 unique combinations of $Q$ and $Q_{\rm p}$ were simulated.
    	
    	\end{tablenotes}
    \end{threeparttable}	
\end{table}

Our computations actually achieve steady state along family $g'$ (Figure~\ref{fig:equilibrium}), with an average value of $e$ much higher than any of the family $g$ orbits, but also not extending to as large $\alpha$ as the far end of family $g'$. Prior work on tidal dissipation in planets and satellites have found stable equilibria, limit cycles, and chaotic phases all to be possible states \citep{2008Meyer}. We tentatively identify family $g'$ as a possible stable equilibrium actually achievable with only small modification due to rotational and tidal bulges and dissipation. The presence of some unsteady behavior is from a combination of limit cycles (where no escape occurs) and chaotic phases (allowing escape). 

We focus attention on how the system achieves equilibrium on family $g'$, and then determine an empirical functional form for the probability of escape. \cite{2009Cassidy} analyzed the problem of a satellite on a variational orbit (family $g$) and found that tidal dissipation in the satellite could considerably heat, even melt, the satellite. Meanwhile, this energy was coming from the orbit of the star around the planet-moon system, not from the orbit of the satellite around the planet. Since the former energy reservoir is enormous, the satellite could be heated on timescales of several billion years with only negligible orbital evolution of the system. However, since the energy does not come at the expense of the orbit of the satellite, and the tides on the planet continue pumping orbital energy into the satellite, this picture does not explain how the satellite can reach a steady state in our computation. 

\subsection{Equilibrium State}
Important to the dynamics of satellites is the evection resonance, which occurs when the pericentre precession rate of the satellite $\dot{\varpi}$ is commensurable with the mean motion of the planet $n_{\rm p}$. This precession is produced by stellar perturbations and by the quadrupole moment of the planetary gravitational potential. For our simulations, stellar perturbations dominate over the planetary quadrupole. However, for satellites at a substantial fraction of the Hill radius, simple methods of orbit averaging are insufficient at analytically approximating the pericentre precession rate. \cite{1969Henon} demonstrated that the evection resonance occurs at $\alpha~{\simeq}~0.41$. Our own numerical simulations confirm this common result, with the resonance argument $\phi = \varpi - \lambda_{\rm p}$ librating near zero, where $\lambda_{\rm p}$ is the orbital longitude of the planet. The dynamical equations for $n$ and $e$ due to the star include only factors of $\sin{\phi}$ \citep{1982Yoder}, so $\phi\simeq0$ implies that the stellar perturbations of these orbital elements are negligible. We can therefore direct our attention to the relevance of tides in the equilibrium state.

Our numerical simulations demonstrate that the satellite remains stable with an excited eccentricity oscillating about some equilibrium value. The stability of this eccentricity is explained by the mutual and counteracting effects of the satellite and planetary tides. We first define the following:
\begin{equation}
    \begin{aligned}[b]
    & A \equiv \frac{9}{2} \frac{k_{\rm p}}{Q_{\rm p}} \frac{m}{m_{\rm p}} \bigg( \frac{R}{a} \bigg)^5, \\
    & B \equiv 7 \frac{ k }{k_{\rm p} } \frac{Q_{\rm p}}{Q} \bigg( \frac{m_{\rm p}}{m}\bigg)^2 \bigg( \frac{R}{R_{\rm p}} \bigg)^5.
    \end{aligned}
\end{equation}
From \cite{1981Yoder}, the equations of motion for $n$ and $e$ due to the mutual tides of both bodies are then
\begin{equation}
    \begin{aligned}[b]
        &\frac{dn}{dt\hfill} = An\big( Be^2 - 1 \big), \\
        &\frac{de^2}{dt\hfill} = - \frac{2}{3} A \big( Be^2 - 1 \big).
    \end{aligned}
\end{equation}
The equation for the tidal change in $e$ holds because tidal dissipation conserves the orbital angular momentum of the satellite $L=mna^2(1-e^2)^{ 1/2}$ for a synchronously rotating satellite in the evection resonance, and all satellites in equilibrium at the end of our simulations have reached the evection resonance and are synchronously rotating. The evection resonance is required for the conservation of $L$ since under this condition, the spin angular momentum of the planet is completely transferred directly into its own orbit and not into the orbit of the satellite. Letting $dn/dt = de/dt = 0$ and solving for eccentricity yields
\begin{equation}
    e^* = B^{-1/2}.
\end{equation}
To validate the stability of this equilibrium, we identify the following Lyapunov function:
\begin{equation}
    V = \big(Be^2-1\big)^2.
\end{equation}
This choice of a Lyapunov function does not include $n$ as an argument, since $e^*$ is independent of $n$ and is an equilibrium value for both $dn/dt$ and $de/dt$. The derivative of the Lyapunov function is
\begin{equation}
    \frac{dV}{dt\hfill}  = -\frac{4}{3} AB \big( Be^2 - 1 \big)^2.
\end{equation}
Since $dV/dt \leq 0$, it follows that $e^*$ is a stable equilibrium. For the most dynamically evolved systems, which are those systems with $\tau > 2\sigma~{\simeq}~0.4$, the value of $e^*$ ranges from $0.08$ to $0.27$. This corresponds closely with the actual eccentricities attained by these systems, which range from $0.05$ to $0.36$. Our analysis demonstrates that evolved systems do reach an equilibrium eccentricity which acts to halt expansion of the satellite's orbit. The mutual tidal effects nullifying each other, the satellite finds itself among the $g'$ orbits, which are known to be stable. Cumulatively, the equilibrium state requires an eccentricity $e=e^*$, synchronous rotation, and the evection resonance, all of which is in accordance with our numerical results.
\subsection{Probability of Escape}
\begin{figure}
    \includegraphics[width=\columnwidth]{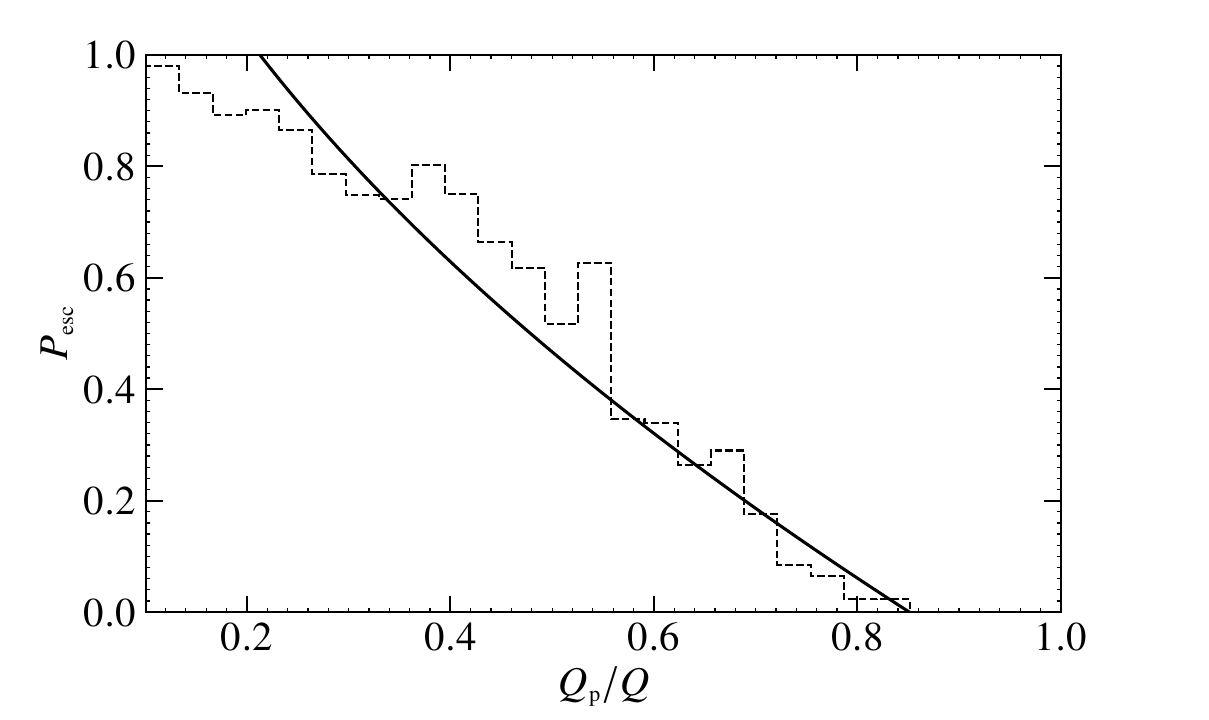}
    \caption{Dotted histogram shows the probability of escape found numerically, while the solid line is the semi-analytic functional form from equation~(\ref{eq:pesc})}
    \label{fig:pfit}
\end{figure}

To determine an expression for the probability of escape, we take the magnitude of the eccentricity excitation due to capture into the evection resonance, and compare this to the maximum stable eccentricity for $g'$ orbits. \cite{1981Yoder} found that capture into a resonance excites the eccentricity by an amount $\Delta e=6^{1/2}e_{\rm c}$, where $e_{\rm c}$ is a critical eccentricity required for transitioning between circulating and librating modes in $\phi$. For us, $e_{\rm c} = e^*$. Meanwhile, the maximum stable eccentricity for $g'$ orbits is $e\simeq0.52$. Assuming $P_{\rm esc}(\Delta e=0.52) = 1$, we propose an ad-hoc function which follows our numerical $P_{\rm esc}$ results: 
\begin{equation}
    P_{\rm esc} = 2 - \frac{0.52}{\Delta e}.
    \label{eq:pesc}
\end{equation}
In Figure~\ref{fig:pfit}, this function is shown as comparable to the numerical probability of escape. However, future work should test various combinations of Love numbers, $Q$-factors, masses, and radii to confirm that equation~\eqref{eq:pesc} holds or identify additional dependence that it misses.

\section{CONCLUSIONS} \label{sec:conclusions}
We have computed the tidal evolution of satellite systems accounting for both tidal dissipation in the planet and in the satellite. If tidal dissipation in an exomoon is comparable or greater than that in their host planet, it can hang out at a substantial fraction of the Hill radius as its host planet's spin rate is decreasing.  In Figure~\ref{fig:simplots} we see that the satellite's orbit comes to an equilibrium in $\alpha$, $e$, and $s$, in which the planet's spin angular momentum is transferred into its own orbit.

\subsection{Application to transiting planets}
A typical planet discovered by Kepler has $R\sim 1.5 R_\oplus$ and orbits at $a\sim0.1~\rm au$. Depending on whether it is a sub-Neptune or a super-Earth, its $Q$ can be as high as $10^4$. Satellites would tend to be smaller, and perhaps rocky, so their $Q$ would tend to be lower. Therefore, this mechanism would likely keep the satellite from escaping, and it would keep them at a large separation from the planet ($a\simeq0.41 \rH$) for hundreds of millions of years. 

Currently, the Sun raises about half as big a tide on the Earth in comparison to the Moon, and the degree of dissipation is a function of how fast this bulge moves through the oceans, which is comparable for each because Earth's spin dominates. Therefore, if the Moon were about $2^{1/3}\simeq1.26$ times further away (i.e. $0.315 \rH$), then the tidal dissipation from the Sun and the Moon would be comparable. At the extreme end of the $g$ family for instance, the Sun's dissipation would be stronger. Our moon would be challenging to detect with even the best photometry \citep{2009Kippingb}, so it is more likely to imagine a bigger, more massive satellite that would in fact dominate the tidal despinning of the planet, relative to the star. 

\subsection{Limitation and future work}

Due to time limitations placed on the usage of computational resources, our simulations did not follow the evolution beyond the quasi-equilibrium in which the moon is synchronous and no longer moving out, but the planet is still spinning relatively rapidly. If the spin rate of the planet keeps decreasing at the same rate as in Simulation 1, then it will become synchronous with the satellite at $\tau\sim2$. The dominant angular momentum transfer would be due to the direct tide the star raises on the planet, which we have not modelled. That continual leakage of angular momentum from the planet-satellite pair would cause the satellite to spiral towards the planet \citep{2023Hansen}. Eventually it could directly impact the planet, if the planet is much less dense, or cause the satellite to be tidally shredded at its Roche limit. After such a shredding episode, the satellite could reconstitute itself as a less massive satellite \citep{2017Hesselbrock}. At some point the satellite mass would no longer dominate the torque on the planet's spin, and the star would be able to spin-synchronise the planet.

Future work could synthesise a population of stars, planets, and moons, at a variety of starting conditions consistent with planet formation, and determine what star-planet separations and planet types would most likely have moons in the tidally-evolved state we have found: $e\simeq0.1$ at $a~{\simeq}~0.41 \rH$. Prior work, for example by \cite{2017Zollinger}, \cite{2020Tokadjian}, and \cite{2023Hansen}, has assumed satellites would spiral outward and escape once they reached a semi-major axis ${\simeq}0.5 r_{\rm H}$, which is what tidal dissipation in satellites can prevent. For some combinations of parameters, we anticipate the lifetimes of moons which would have otherwise escaped would be prolonged, increasing their chance of being detected. { Tidal dissipation may cause such significant heating in these exomoons, that they would shine brightly in the infrared \citep{2013Peters-Limbach} and  produce observable, volcanic signatures of sodium and potassium \citep{2019Oza}, making such exomoons even more detectable.} We are encouraged that the exomoon-hunting era is not over with the results of Kepler; we anticipate moons surviving around planets with periods shorter than the Earth. Hence, TESS \citep{2015Ricker}, JWST \citep{2006Gardner}, PLATO \citep{2022Rauer}, and others have a role to play in the hunt for exomoons. 

\section*{ACKNOWLEDGEMENTS}
Code from \cite{2002MardlingLin} was kindly provided by R. Mardling. We thank R. Mardling, B. Quarles, P. Arras, D. Kipping, A. Oza, and an anonymous referee for helpful suggestions for revision. We thank the Research Computing Center at the University of Chicago for use of the Midway2 cluster, where our simulations were computed. This work was supported by the University of Chicago Jeff Metcalf Fellowship Grant. 

\section*{DATA AVAILABILITY}
N-body simulation data may be downloaded at the following DOI: 10.5281/zenodo.8306687.



\bibliographystyle{mnras}
\bibliography{example} 






\bsp	
\label{lastpage}
\end{document}